\documentclass[%
 reprint,
superscriptaddress,
amsmath,amssymb,
aps,
prstab,
]{revtex4-2}

\usepackage{graphicx}
\usepackage{dcolumn}
\usepackage{bm}

\begin{document}

\title{Field, frequency and temperature dependence of the surface resistance of nitrogen diffused niobium superconducting radio frequency cavities}

\author{P. Dhakal}
 \email{dhakal@jlab.org}
 \affiliation{Thomas Jefferson National Accelerator Facility, Newport News, VA 23606, USA}
 \affiliation{Center for Accelerator Science, Department of Physics, Old Dominion University, \\Norfolk, Virginia 23529, USA}
 \author{B.D. Khanal}
  \affiliation{Center for Accelerator Science, Department of Physics, Old Dominion University, \\Norfolk, Virginia 23529, USA}
\author{A. Gurevich}
  \affiliation{Center for Accelerator Science, Department of Physics, Old Dominion University, \\Norfolk, Virginia 23529, USA}
 \author{G. Ciovati}
 \affiliation{Thomas Jefferson National Accelerator Facility, Newport News, VA 23606, USA}
 \affiliation{Center for Accelerator Science, Department of Physics, Old Dominion University, \\Norfolk, Virginia 23529, USA}
\date{\today}

\begin{abstract}
We report the RF performance of several single-cell superconducting radio-frequency cavities subjected to low temperature heat treatment in nitrogen environment. The cavities were treated at temperature 120 - 165 $^{\circ}$C for an extended period of time (24 - 48 hours) either in high vacuum or in a low partial pressure of ultra-pure nitrogen. The improvement in $Q_0$ with a Q-rise was observed when nitrogen gas was injected at $\sim$300 $^{\circ} $C during the cavity cooldown from 800 $^{\circ}$C and held at 165 $^{\circ}$C, without any degradation in accelerating gradient over the baseline performance. The treatment was applied to several elliptical cavities with frequency ranging from 0.75 GHz to 3.0 GHz, showing an improved quality factor as a result of low temperature nitrogen treatments. The Q-rise feature is  similar to that achieved by nitrogen alloying Nb cavities at higher temperature, followed by material removal by electropolishing. The surface modification was confirmed by the change in electronic mean free path and tuned with the temperature and duration of heat treatment. The decrease of the temperature-dependent surface resistance with increasing RF field, resulting in a Q-rise, becomes stronger with increasing frequency and decreasing temperature. The data suggest a crossover frequency of $\sim 0.95$~GHz above which the Q-rise phenomenon occurs at 2~K. Some of these results can be explained qualitatively with an existing model of intrinsic field-dependence of the surface resistance with both equilibrium and nonequilibrium quasiparticle distribution functions. The change in the Q-slope below 0.95 GHz may result from masking contribution of trapped magnetic flux to the residual surface resistance.
\end{abstract}

\maketitle


\section{Introduction}
Recent advances in the processing of bulk superconducting radio frequency (SRF) niobium cavities via interior surface impurity diffusion have resulted in significant improvements in their quality factor ($Q_0$). The motivation for the development of these processes is to reduce the cryogenic operating cost of current and future accelerators while providing reliable operation \cite{Dhakalipac,Dhakalprab,AnnaSUST1,DhakalASC,DhakalPO}. Most recently, efforts have been made to preserve high accelerating gradients, $E_{acc}$, while also increasing the quality factor of SRF cavities \cite{AnnaSUST2,Koufa,Dhakalinf}. In the literature, these cavity processing recipes are referred to "nitrogen infusion” cavity processing recipes where, cavities were heat treated at 800 $^{\circ}$C for 3 hours, then the furnace temperature is reduced to 120 - 200 $^{\circ}$C and nitrogen is introduced into the furnace at a partial pressure of $\sim$ 25 mTorr for $\sim$ 48 hours. This process has shown an improvement in $Q_0$ over the baseline measurements, without the need for post-annealing chemical etching with no significant reduction in quenched field. Even though diffusion of the nitrogen into the bulk of the SRF cavity is limited in depth ($\sim$ 50 nm) at these low temperatures (120 - 200 $^{\circ}$C), the introduction of nitrogen is sufficient to modify the cavity surface within the RF penetration depth as seen from RF results, which are similar to those previously reported for high-temperature nitrogen treated cavities. Surface analysis done on the sample coupons treated with cavities showed a complex NbN$_{1-x}$O$_x$ on the surface of SRF cavities, which may be responsible for the Q-rise in those cavities \cite{Dhakalinf}. The absence of post furnace chemical treatment shows a clear benefit in reducing processing steps as well as keeping higher gradient with high $Q_0$ values.  The field dependence of the surface resistance calculated based on the model which extends the Bardeen-Cooper-Schrieffer (BCS) surface resistance to high rf field qualitatively explained the experimental results \cite{Dhakalinf,AlexSUST}.

More recently, the medium temperature heat treatments of the SRF cavities have been investigated  to explore the mechanism of impurities diffusion as well as exploration of new dopants. The  heat treatment of SRF cavities in the  range of 300 - 400 $^{\circ}$C in vacuum resulted in the increase in quality factors with reduced RF field similar to those previously observed in impurity diffused SRF cavities \cite{sam,Ito,Eric,Zhou}. In this work we present the results of RF tests on single cell elliptical cavities with different sizes treated in nitrogen at low temperatures to explore the frequency and field dependencies of the quality factor at different temperatures. The experimental data are interpreted using different theoretical models.

\section{Cavity Surface Preparations}
Several single cell cavities of frequencies 0.75 - 3.0 GHz were selected for the current study. All of the cavities are fabricated from high-purity (residual resistivity ratio $>$ 250), fine grain (ASTM $>$5) niobium. A summary of the cavities' electromagnetic parameters are listed in Table \ref{table1}. Such parameters include the ratio of the peak surface electric and magnetic fields divided by $E_{acc}$, $E_p/E_{acc}$ and $B_p/E_{acc}$, respectively, $k = \frac{1}{L}\sqrt{\frac{R}{Q}}$, where $L$ is the cavity active length and $R/Q$ is the shunt impedance, and the geometry factor, $G$. Prior to the RF measurements reported in this manuscript, these cavities went through several cycles of heat treatments, chemical polishing and nitrogen diffusion. The baseline RF measurements reflect the surface reset via electropolishing by removing $\sim$ 20 - 40 $\mu$m from the inner cavity surface.

\begin{table*}
\caption{\label{table1}
Summary of electromagnetic parameters for each of the cavities used in this study.}
\centering
\begin{tabular}{cccccc}
\textrm{Cavity Name}&
\textrm{Frequency (GHz)}&
\textrm{$E_p/E_{acc}$}&
\textrm{$B_p/E_{acc}$[mT/(MV/m)]}&
\textrm{$k$ $\left (\sqrt{\Omega}/\text{m} \right )$}&
\textrm{G ($\Omega $)}\\
\hline
750 MHz & 0.75 & 2.24 & 4.18 & 50.9 & 276.3  \\
RDT-06 & 1.3 & 1.85 & 4.23 & 89.8 & 277.8   \\
RDL-02 & 1.5 & 1.99 & 4.18 & 115.7 & 277.2  \\
FH3C & 3.0 & 1.83  & 4.23 & 207.4 & 277.9  \\	
\hline
\end{tabular}
\end{table*}

\begin{figure}[htb]
\includegraphics*[width=85mm]{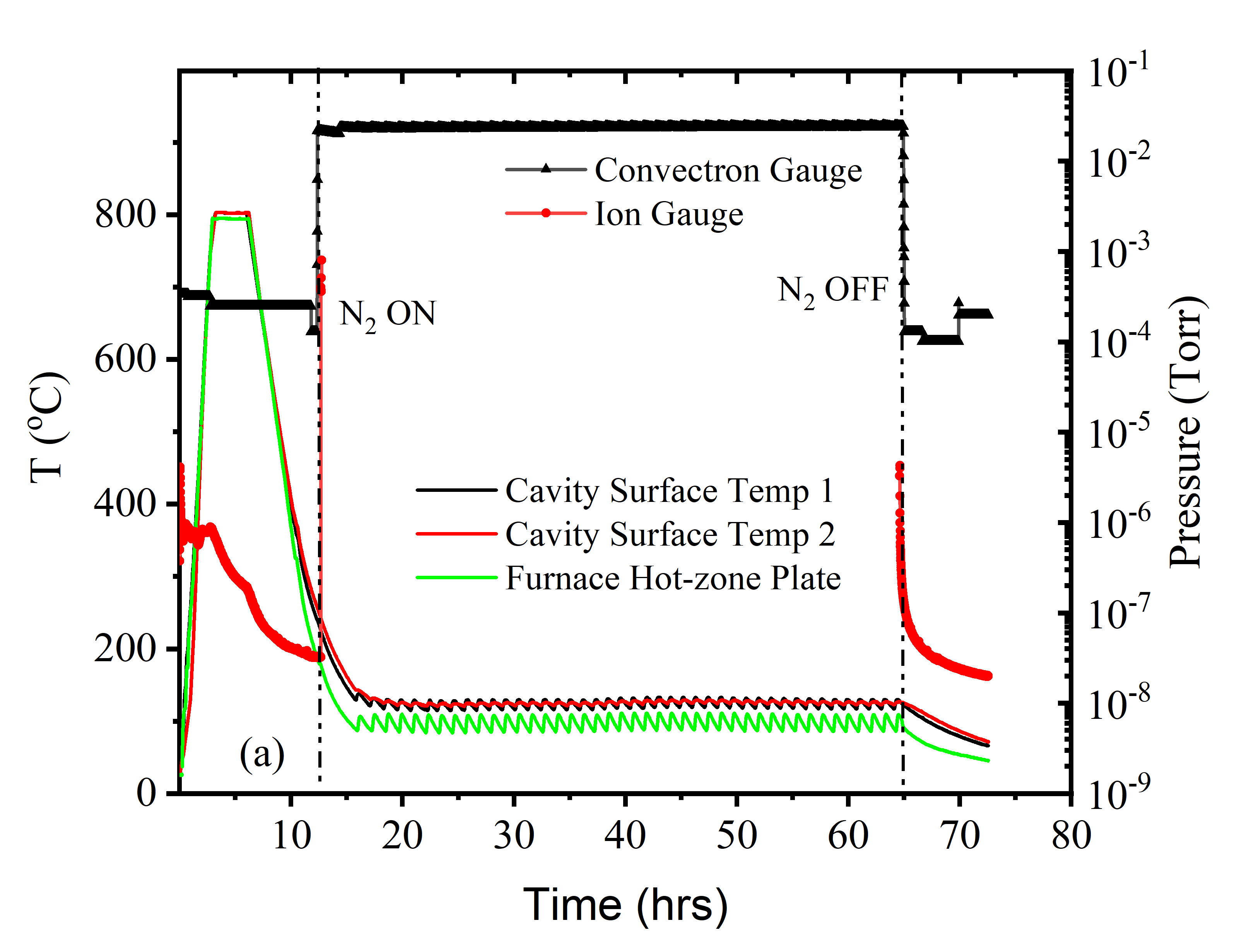}
\caption{\label{fig:furnacerun} Typical heat treatment cycle. Ultra-pure N$_2$ gas can be injected into the furnace at different temperatures during the cool down. Notice that the temperature measured on the cavity surface is higher than that measured in the furnace hot-zone.}
\end{figure}

Before the heat treatment, the cavities were high pressure rinsed and then dried in an ISO 4 cleanroom. While in the cleanroom, special caps made from niobium foils were placed to cover the cavity flange openings \cite{dhakalsrf19}. The cavity was then transported to the furnace in a clean sealed plastic bag. The vacuum heat treatment procedure started with the 800 $^{\circ}$C/3h hydrogen degassing step followed by lowering the temperature of the furnace to (120 - 165 $^{\circ}$C) range. The furnace is continuously pumped during the cooldown process. In temperature range $\sim$ 120 - 300 $^{\circ}$C at which the total pressure (also corresponds to nitrogen partial pressure) increased to $\sim$25 mTorr by introducing high purity nitrogen. Such pressure was maintained without active pumping of the furnace enclosure. Once the temperature has fallen to the desired value (120 - 165 $^{\circ}$C), which takes about 2 hours, the temperature was held for 24 - 48  hours without active pumping. Figure \ref{fig:furnacerun} shows the temperature and pressure profile during heat treatment. The treatment referred to as low temperature baking (LTB) at 120 $^{\circ}C$ was done in ultra-high vacuum (UHV) condition while the cavity is actively pumping on a vertical test stand, immediately before insertion in the vertical cryostat.

\section{Cavity Test Results}
\subsection{RF Test Results}
Standard procedures were followed to clean the cavity surface in preparation for an rf test: degreasing in ultra-pure water with a detergent and ultrasonic agitation, high pressure rinsing with ultra-pure water, drying in the ISO 4/5 cleanroom, assembly of flanges with RF feedthroughs and pump out ports and evacuation. The cavity was inserted in a vertical cryostat and cooled to 4.2 K with liquid helium using the standard Jefferson Lab cooldown procedure in a residual magnetic field of $< 2$~mG along the cavity axis. This procedure results in a temperature difference between the two irises $\Delta T > 4$~K when the equator temperature crosses the superconducting transition temperature ($\sim$ 9.25 K), which provides good flux expulsion conditions \cite{dhakal20}. Most of the rf measurements were done at a He bath temperature $T = 2.0$~K, acquiring $Q_0(E_{acc})$ curves and in some instances $Q_0(T) = G/R_s(T)$ from 4.2 -1.6 K were obtained at a constant peak surface magnetic field in order to extract the residual resistance, $R_i$. 

\subsubsection{750 MHz}

\begin{figure}[htb]
\includegraphics*[width=85mm]{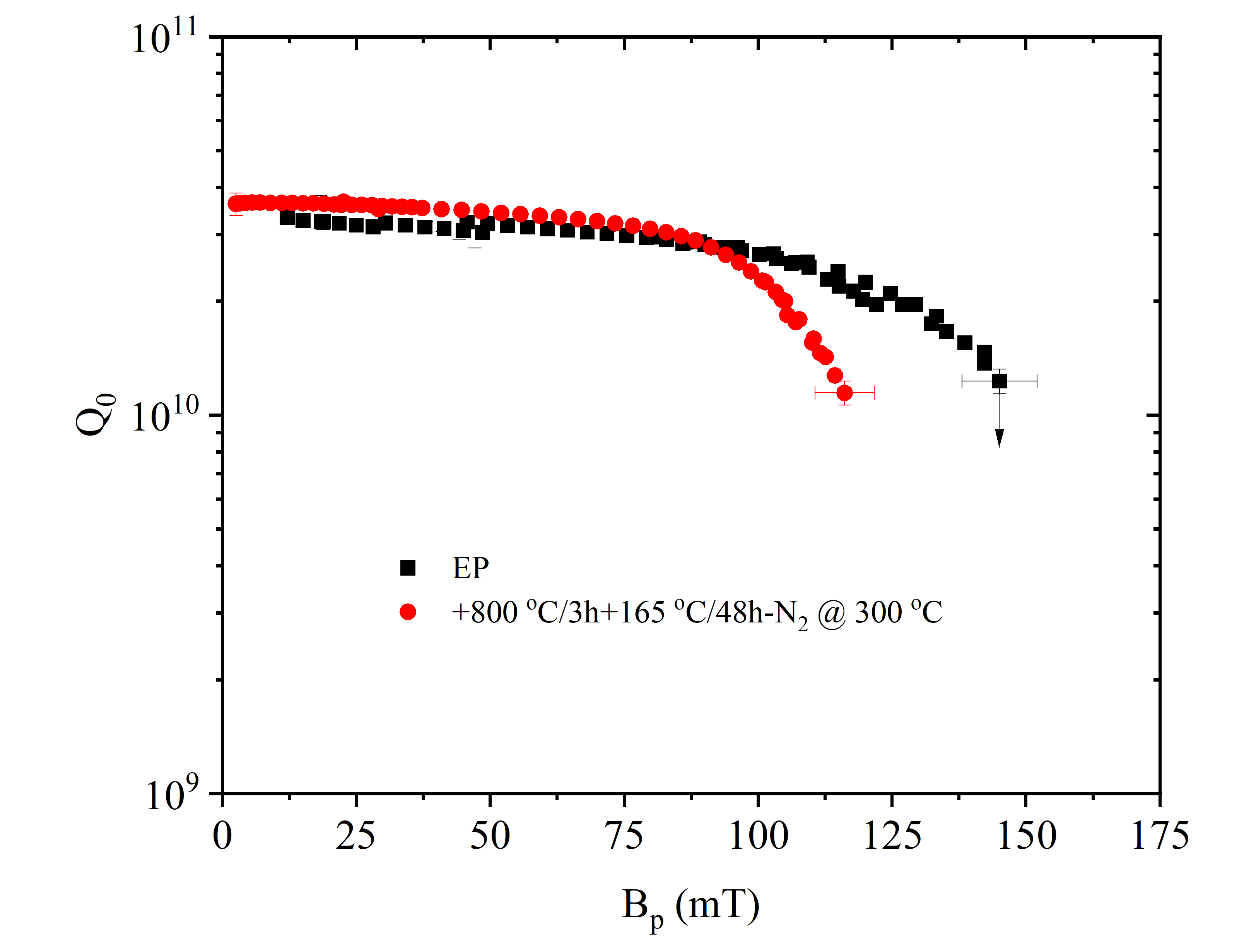}
\caption{\label{fig:750} $Q_0(B_p)$ at 2.0 K for cavity 750 MHz single cell. The arrow represents that the cavity was limited by quench. The test after annealing was limited by field emission.}
\end{figure}

The RF test summary of $Q_0$ vs. $B_p$ measured at 2.0 K for cavity 750 MHz is shown in Fig. \ref{fig:750}. The cavity was previously treated with buffered chemical polishing \cite{750}. The baseline measurement was done after the cavity was subjected to $\sim$ 20 $\mu$m EP. The cavity was limited by quench at $B_p$= 145$\pm$3 mT. The cavity was then subjected to high temperature heat treatment at 800 $^{\circ}$C/3h followed by the nitrogen injection into the furnace at 300 $^{\circ}$C. The temperature of the cavity was held at 165 $^{\circ}$C for 48 hours. After annealing, the cavity RF performance was limited by strong multipacting at $B_p$= 50 $\pm$ 5 mT. This could have resulted from a small shape deformation of the cavity shape after evacuation, as a result of a reduced yield strength of the Nb after the heat treatment at 800 $^{\circ}$C. The cavity was mechanically stretched back to its initial frequency and it was restrained from deformation during the evacuation, in preparation for the subsequent RF test at 2~K, after high-pressure water rinsing. The strong multipacting was no longer observed but field emission started at $B_p \sim $ 60 mT during the RF test. Helium processing \cite{Padamsee} mitigated the field emission, which still limited the cavity to $B_p \sim 115$~mT. A small increase in quality factor at low rf field was observed, with no apparent Q-rise, after the N-infusion treatment.
The Q-rise has been associated with a decrease of the BCS surface resistance with increasing rf field in both 1.3-1.5 GHz Ti-and N-diffused cavities \cite{GigiAPL14, AnnaSUST1}. In order to verify the absence of any decrease of $R_{\text{BCS}}(B_p)$, $R_s(T)$ was measured between 1.7 - 4~K for different $B_p$-values, ranging between 9 - 42 mT, as shown in Fig.~4(a). The data were fit with the following generic form of thermally-activated surface resistance at $T\ll T_c$, for each $B_p$-value, with the same method described in Ref. \cite{GigiAPL14}:
\begin{equation}
R_s(T_s)=\frac{A}{T_s}e^{-U/k_BT_s}+R_i,
\label{eq:one}
\end{equation}
where $T_s$ is the temperature of the cavity inner surface and $k_B$ is the Boltzmann constant. A self-consistent calculation of $T_s$ becomes important above 25 mT, in order to clearly separate RF heating from the intrinsic field dependence of $R_s(B_p)$. Figure ~4(b-d) shows the values of the fit parameters \textit{A}, \textit{U}, $R_i$ as a function of $B_p$, showing an increase of the pre-exponential factor \textit{A} with increasing RF field, unlike what was measured in N and Ti-doped cavities above 1~GHz.

\begin{figure}[htb]
\includegraphics*[width=85mm]{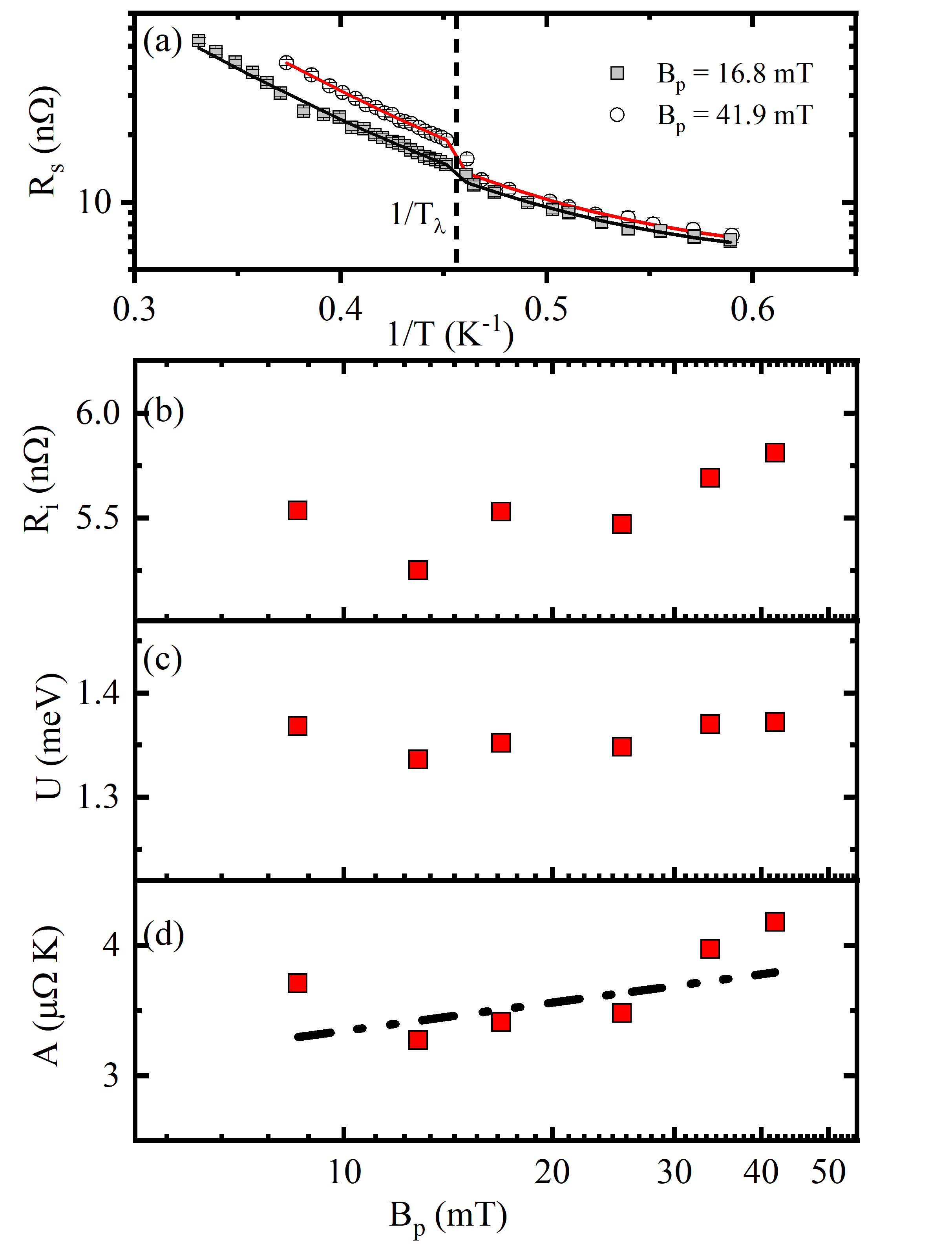}
\caption{\label{fig:750RST} $R_s(T)$ measured at different $B_p$-values 16.8 and 41.9~mT (a) and fit parameters values from least-square fits with Eq.~(\ref{eq:one}). Solid lines in (a) are from fits with Eq.~(\ref{eq:one}) and the dash-dot line in (d) is a linear fit to the data. The vertical line in (a) is at $1/T_{\lambda}$=1/2.17 K.}
\end{figure}

\subsubsection{RDT-06}

\begin{figure}[htb]
\includegraphics*[width=85mm]{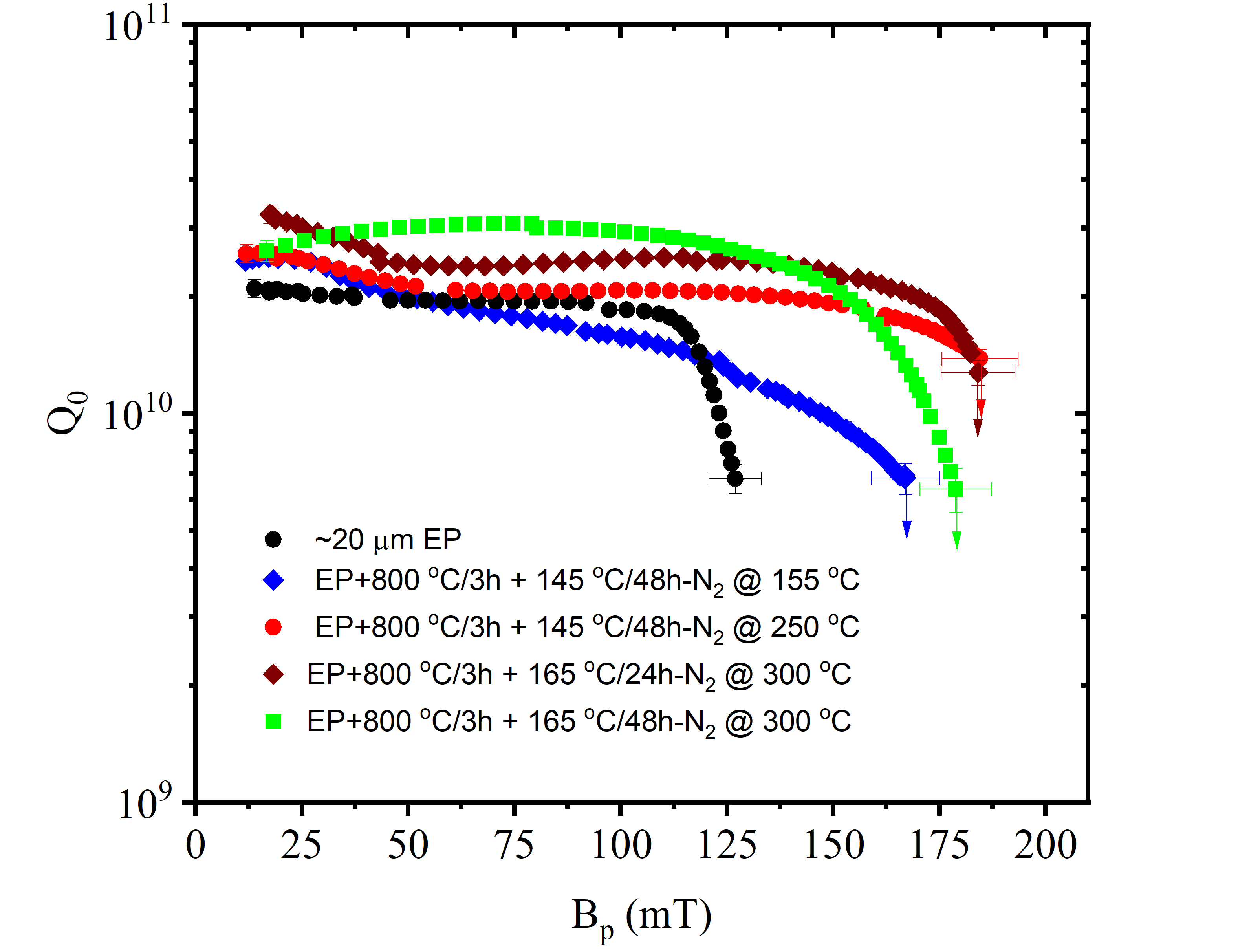}
\caption{\label{fig:RDT-6} $Q_0(B_p)$ at 2.0 K for cavity RDT-6. The arrows indicate the field above which the cavity quenched. There was no field emission in any of the tests.}
\end{figure}

The RF test summary of 1.3 GHz cavity labeled RDT-6 is shown in Fig. \ref{fig:RDT-6}. The cavity was treated at different temperature while varying the nitrogen injection temperature. The baseline measurement was done after the cavity was subjected to $\sim$20 $\mu$m EP. The cavity was limited at $B_p$= 127$\pm$6 mT by the high field Q-slope. It is to be noted that after each of the annealing listed in Fig. \ref{fig:RDT-6}, the RF surface of cavity was reset by $\sim$ 20 $\mu$m EP.  The cavity was then subjected to high temperature heat treatment at 800 $^{\circ}$C/3h followed by the nitrogen injection into the furnace at 155 $^{\circ}$C. The temperature of the cavity was held at 155 $^{\circ}$C for 48 hours. The RF performance showed the increase in peak RF field to $B_p$ = 167$\pm$8 mT with $Q_0$ = 6.8 $\times 10^9$ before the cavity quench. The low field $Q_0$ increases but it decreases with increasing RF field. After the cavity reset by EP, it was again treated at 800 $^{\circ}$C/3h followed by the nitrogen injection into the furnace at 250 $^{\circ}$C. The temperature of the cavity was then held at 155 $^{\circ}$C for 48 hours. The RF performance showed the increase in $Q_0$ above 90 mT with $Q_0 \sim 1.4\times 10^{10}$ at quench field of $B_p$ = 185$\pm$9 mT. The next heat treatment was done at nitrogen injection into the furnace at 300 $^{\circ}$C and the cavity was held at 165 $^{\circ}$C for 48 hours. The RF performance showed the $Q_0$ increase with RF field, similar to those observed in cavities with highest $Q_0$ $\sim$ 3$\times10^{10}$ at $B_p$ $\sim$ 90 mT. The $Q_0$ at high field showed the board Q-slope before it quenched at $B_p$ = 179$\pm$8 mT with $Q_0 \sim 6.4\times 10^{9}$. The final heat treatment was done at the same temperature as the earlier one, but the cavity was held at 165 $^{\circ}$C only for 24 hours. Again, an overall increase in $Q_0$ was observed before the cavity quenched at $B_p$ = 185$\pm$8 mT with $Q_0 \sim$ 1.3$\times 10^{10}$. The maximum achievable RF field didn't degrade as result of nitrogen infusion. The results show that $Q_0$ can be tuned depending on the temperature of the nitrogen injection into the furnace, duration, and temperature of the cavity during the heat treatment. 
\subsubsection{RDL-02}
\begin{figure}[htb]
\includegraphics*[width=85mm]{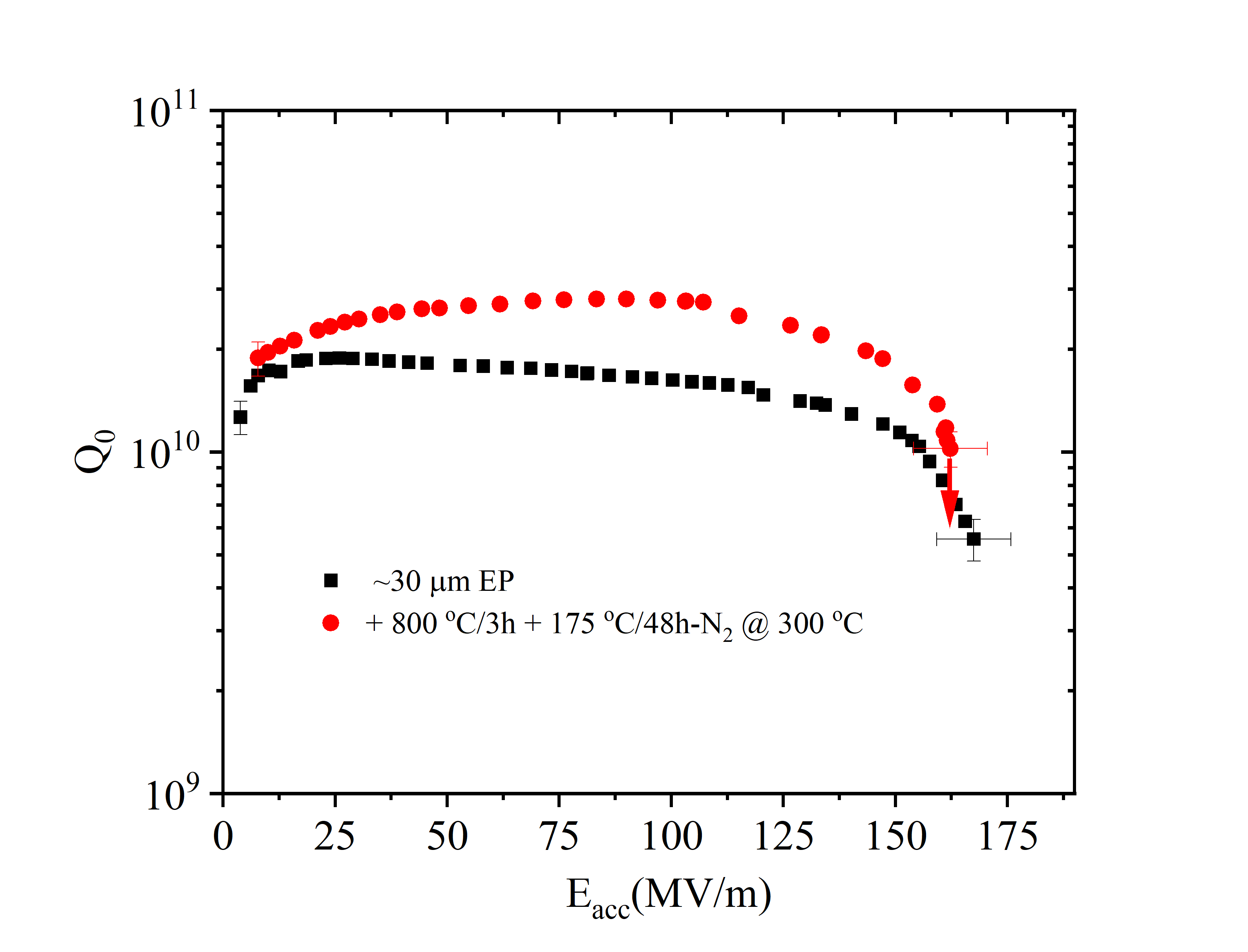}
\caption{\label{fig:RDL-02} $Q_0(B_p)$ at 2.0 K for 1.3 GHz cavity GHz RDL-02. The arrows indicate the field above which the cavity quenched. There was no field emission in any of the tests.}
\end{figure}

The 1.5 GHz cavity labeled RDL-02 was treated with the profile in the way similar to that shown in Fig. \ref{fig:furnacerun}, where the nitrogen gas was injected at 300 $^{\circ}$C. The baseline measurement was limited by high the field Q-slope at $B_p$ = 168$\pm$5 mT. After the nitrogen infusion the cavity reached   $B_p$ = 162$\pm$5 mT and limited by quench as shown in Fig. \ref{fig:RDL-02}. The $Q_o(B_p)$ curves shows the Q-rise phenomenon with increasing $Q_0$ with peak RF magnetic field up to 2.75$\times 10^{10}$ at $\sim$ 107 mT corresponding to accelerating gradient of 26 MV/m. The Q-rise phenomenon is similar to those observed in high temperature nitrogen treated cavities. The details of analysis with cavity treated at different temperature was already presented in Ref. \cite{Dhakalinf}. 
 \subsubsection{FH3C}
\begin{figure}[htb]
 \includegraphics*[width=85mm]{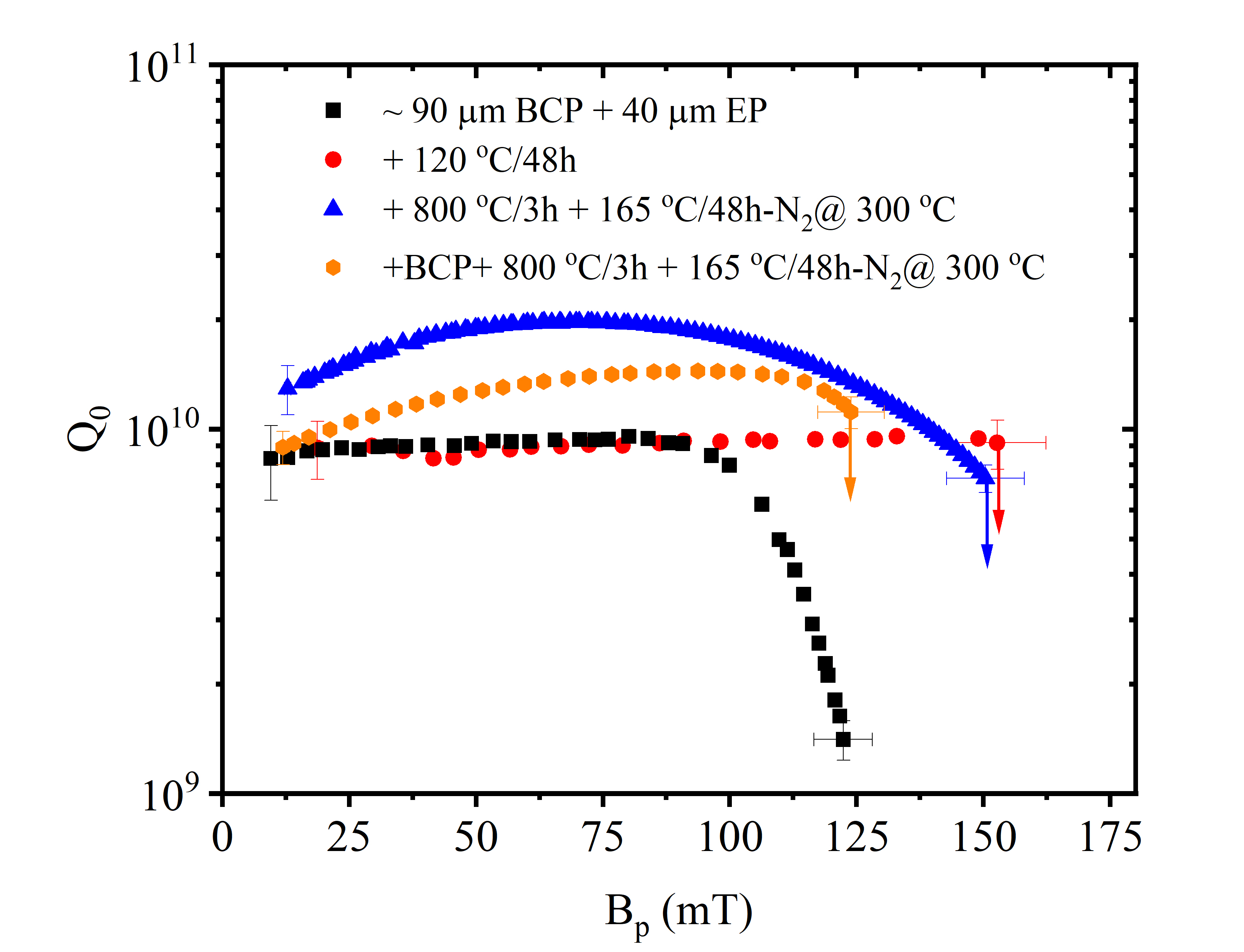}
 \caption{\label{fig:F3HC} $Q_0(B_p)$ at 2.0 K for 3.0 GHz cavity FH3C. The arrows indicate the field above which the cavity quenched. There was no field emission in any of the tests.}
 \end{figure}

 The summary  of the RF tests of 3.0 GHz cavity labeled FH3C is shown in Fig. \ref{fig:F3HC}. The baseline test was done after the rf surface was treated with $\sim$ 40 $\mu$m EP \cite{ciovati2018}. The cavity was limited by high field Q-slope at $B_p$ $\sim$ 125$\pm$5 mT with $Q_0 \sim 1.4\times10^{9}$. The cavity was then in-situ baked on vertical test stand at 120 $^{\circ}C$ for 48 hours while the cavity was actively pumped. As expected, the high field Q-slope is eliminated  and cavity reached $B_p$ = 152$\pm$6 mT with $Q_0 \sim 9.2\times10^{9}$. The cavity was then subjected to additional heat treatment at 800 $^{\circ}$C/3h followed by the nitrogen injection into the furnace at 300 $^{\circ}$C during the cooldown and the cavity temperature was then held at 165 $^{\circ}$C for 48 hours. The RF performance showed the increase in $Q_0$ with rf field up to $\sim$ 75 mT and it was limited by quench at $B_p$ = 150$\pm$6 mT with $Q_0 \sim 7.4\times10^{9}$. The maximum in $Q_0$ = 2$\times 10^{10}$ was observed at a corresponding accelerating gradient of $\sim$ 18 MV/m.
 
\begin{figure}[htb]
 \includegraphics*[width=85mm]{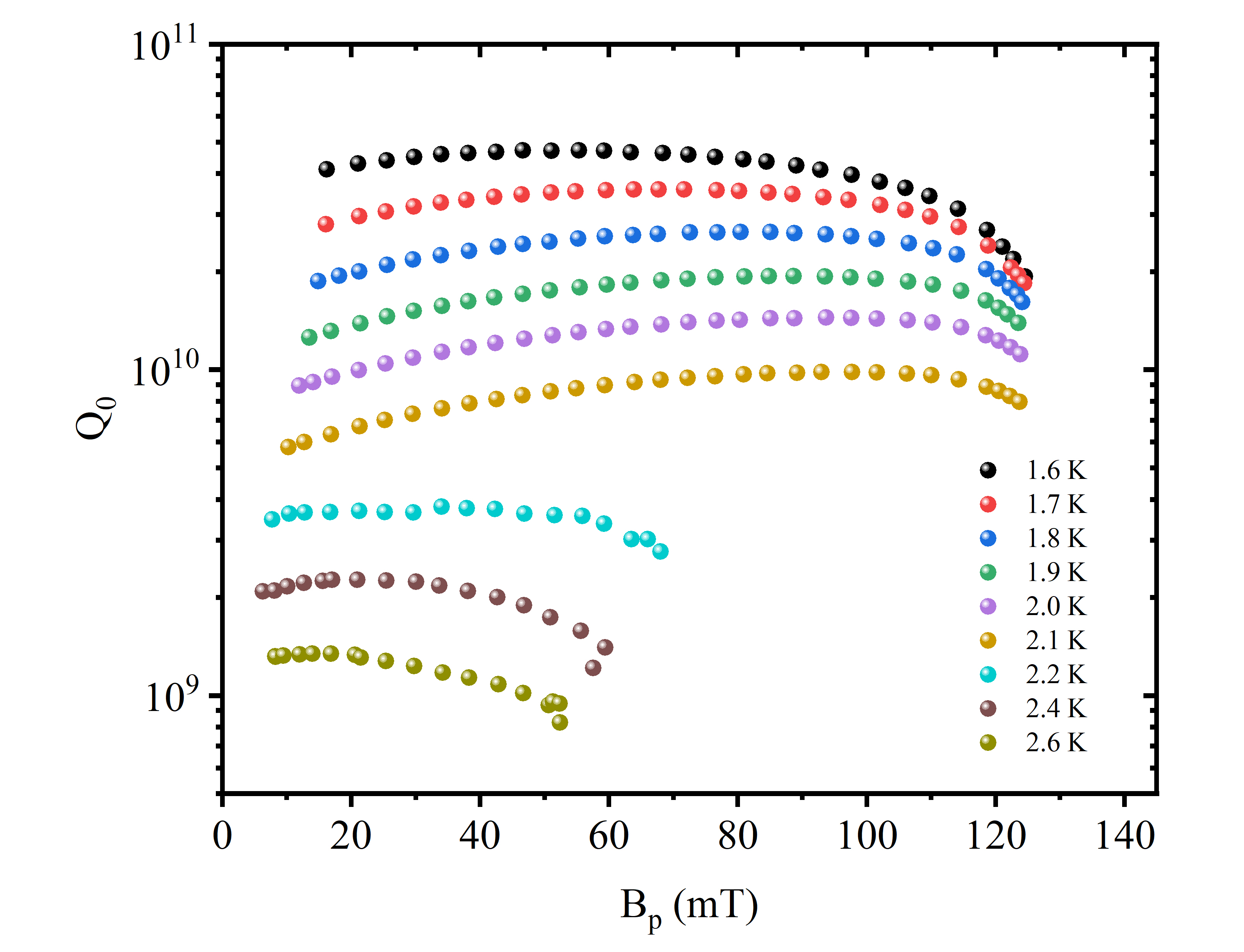}
 \caption{\label{fig:FH3C_new} $Q_0(B_p)$ at 2.6 - 1.6 K for 3.0 GHz cavity F3HC. The tests  at $< 2.1$~K were limited by quench, whereas tests $\ge$ 2.2~K were limited by instability due to heating.}
 \end{figure}

\begin{figure}[htb]
 \includegraphics*[width=85mm]{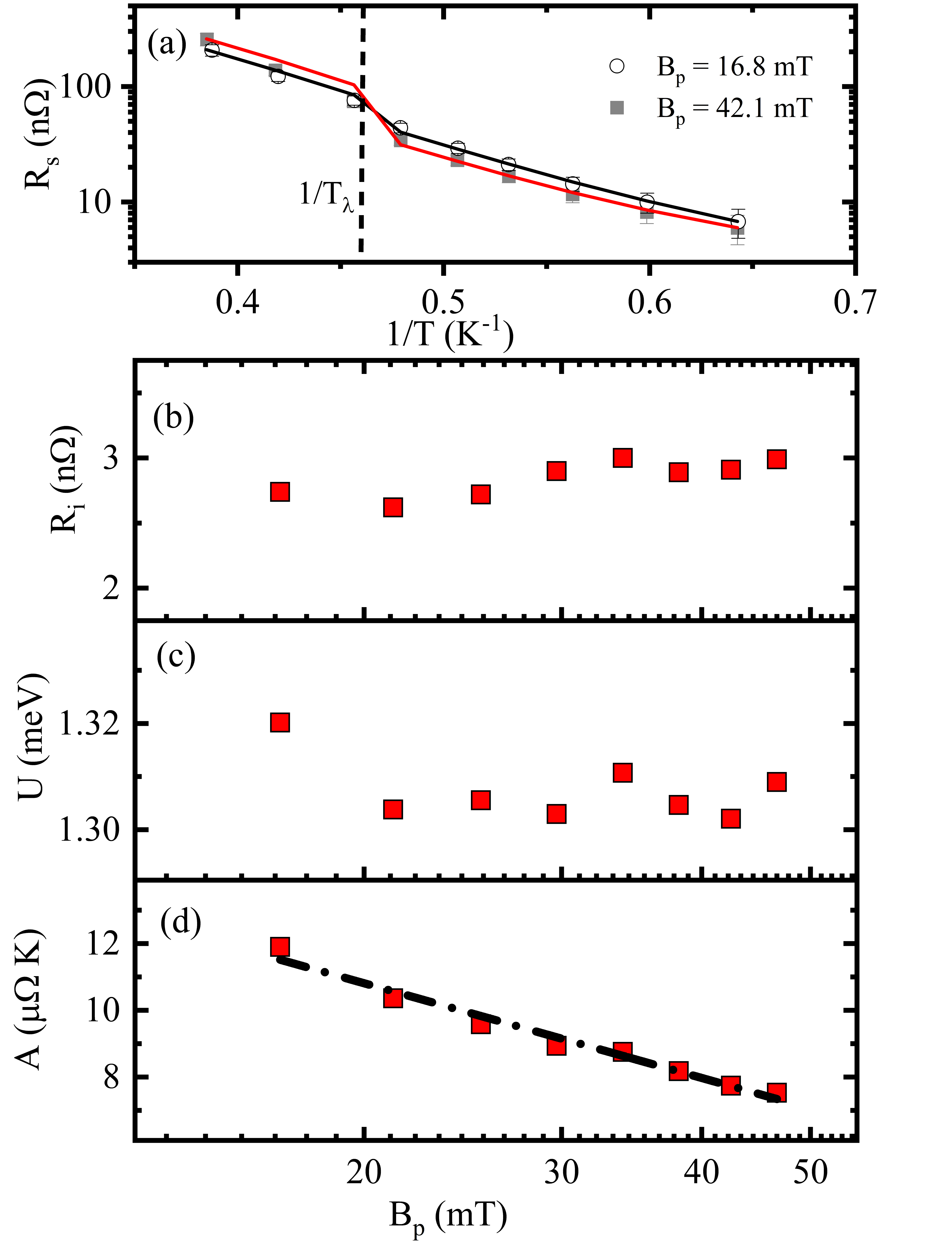}
 \caption{\label{fig:3RS_T}(a) $R_s(T)$ measured at different $B_p$-values 16.8  and 42.3 mT and (c-d) fit parameters values from least-square fits with Eq. (2). Solid lines in (a) are from the fits with Eq.~\ref{eq:one} and the dash-dot line in (d) is a linear fit to the data. The vertical line in (a) is at $1/T_{\lambda}$=1/2.17 K.}
 \end{figure}

The cavity was stored  for $\sim$ 2 years exposed to air and retest resulted in poor performance with high residual resistance. The cavity surface was reset with $\sim$ 10 $\mu$m BCP and the treatment was performed. At this time, the quality factor as a function of RF field were measured at different temperature from 2.6 - 1.6 K as shown in Fig. \ref{fig:FH3C_new}. The cavity is limited by quench at $B_p$ = 124 mT when RF test was performed in super-fluid helium, whereas at higher temperature the cavity was limited by heating likely due to the inefficient heat transfer to liquid helium bath. The temperature dependence of surface resistance was fitted using Eq. (\ref{eq:one}) to extract the residual resistance $R_i$ and material parameters $U$ and $A$ that contribute to the BCS resistance. Figure \ref{fig:3RS_T} shows the values of the fit parameters $A$, $U$ and $R_i$ as a function of $B_p$, showing the decrease in $A$ with increasing RF field. The logarithmic dependence of $A(B_p)$ is consistent with the theoretical model explaining the Q-rise phenomenon \cite{AlexSUST, GigiAPL14}. 
 
\subsection{RF penetration depth}
\begin{figure}[htb]
\includegraphics*[width=85mm]{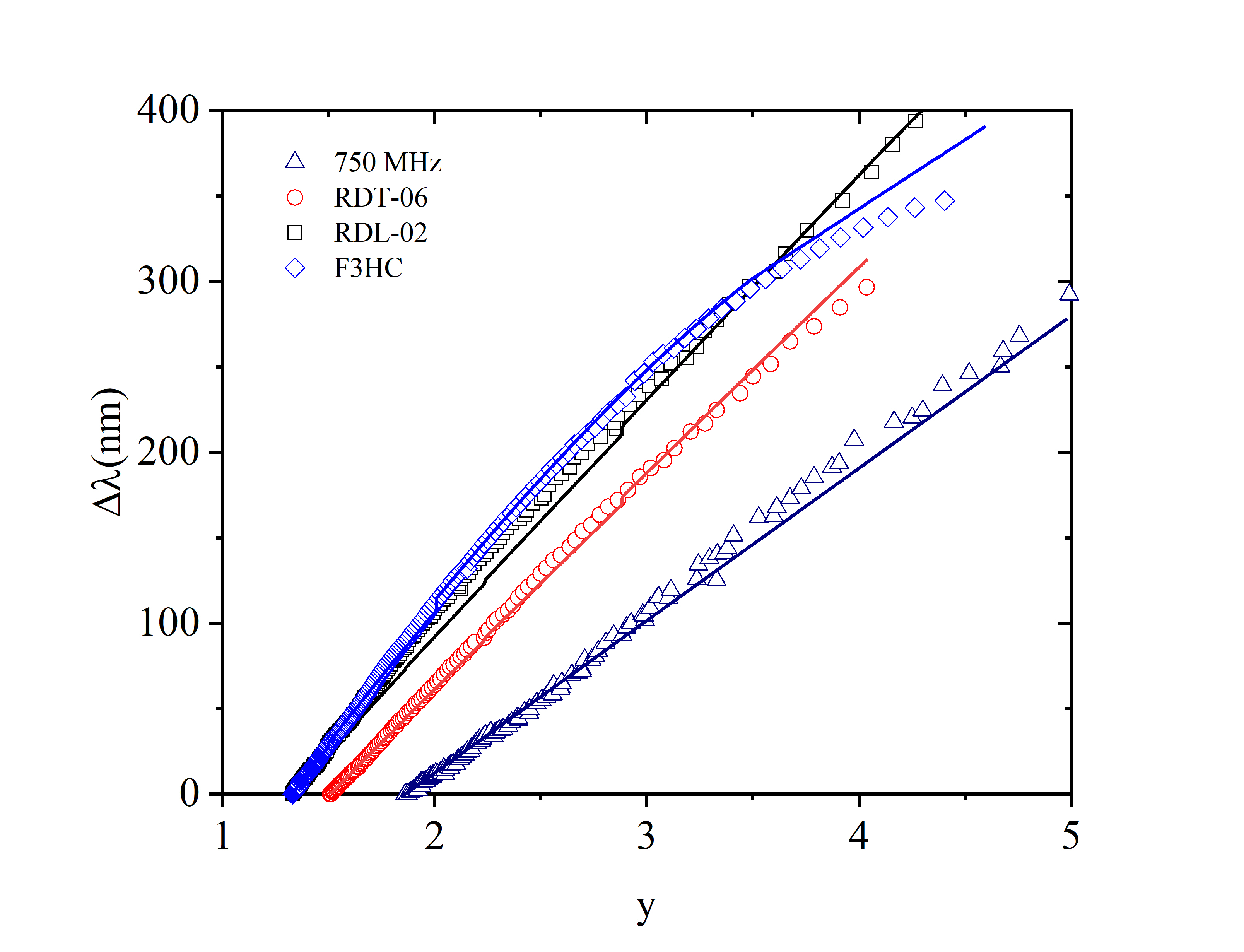}
\caption{\label{fig:ALL_DLL} The change in penetration depth as a function of reduced temperature parameter $y=\sqrt{(1-(T/T_c)^4)}$ for all cavities that underwent the same heat treatment process (EP+ 800 $^\circ$C /3h + 165 $^\circ$C/48h $N_2$@ 300 $^\circ$C . The solid lines are fit with M-B theory.}
\end{figure}

In order to obtain information about the  mean free path near the RF surface, we measured the resonant frequency and quality factor while warming up the cavities from $\sim$ 5 K to above transition temperature ($>$ 9.3 K) using a vector-network analyzer and low-noise rf amplifiers, from which $R_s(T)$ and the change in resonant frequency can be extracted. The frequency shift can be translated into a change in penetration depth according to 
  \begin{equation}
 \Delta \lambda=  \dfrac{G}{\pi \mu_0 f^2}\Delta f
\end{equation}
where $G$ is the geometric factor of the cavity and $f$ is the resonant frequency. Using the Casimir-Gorter relation \cite{CG}, we can obtain $\lambda_0$, which is the penetration depth at 0 K as:
\begin{equation}
 \Delta \lambda= \lambda(T)-\lambda_0= \dfrac{\lambda_0}{\sqrt{(1-T/T_c)^4}} - \lambda_0
\end{equation}
In the Pippard-limit \cite{pippard}, $\lambda_0$ is directly related to the mean free path of the quasi-particles as:
\begin{equation}
 \lambda_0= \lambda_L \sqrt{1+\frac{\pi \xi_0}{2l}}.
\end{equation}
The data were fitted with the  Mattis-Bardeen (M-B) theory \cite{MB} to extract the mean free path and summarized in Table \ref{table2}. The lower mean free path $l < \xi_0/2$, shows that the Nb is in dirty limit due to the low temperature heat treatment in nitrogen environment.  Figure \ref{fig:ALL_DLL} shows the measured  $\Delta \lambda$  vs. the reduced temperature parameter, y = $1/\sqrt{(1-T/T_c)^4}$ for all cavities under study for the same treatment of 800~$^{\circ}$C/3h and the nitrogen injection into the furnace at 300~$^{\circ}$C followed by the holding temperature of 165~$^{\circ}$C/48 h. 


\begin{table*}
\caption{\label{table2}
Results of $R_s(T)$ and $\Delta\lambda(T)$ fits. The mean free path, $l$, obtained from the fit of $\Delta \lambda$ (T) in temperature range (8.0 - 9.25 K) for different treatment and $\Delta/k_BT_c$ obtained from the fit of $R_s(T)$ from temperature 4.3 - 1.6 K for different treatment when data were available. The transition temperature $T_c$ measured during the cavity warm up and kept constant during the fits.}
\centering
\begin{tabular}{ccccccc}
\textrm{Cavity Name}&
\textrm{Frequency (GHz)}&
\textrm{Treatment}&
\textrm{$\Delta /k_BT_c$}&
\textrm{$R_i (n \Omega $) }&
\textrm{$l$ (nm)} &
\textrm{$T_c$ (K)}\\
\hline
750 MHz & 0.75 & $\sim$ 25 $\mu$m EP & $1.78 \pm 0.02$ & $6.1 \pm 0.2$ & 390 $\pm 10$ & 9.30$\pm0.04$ \\
		&  & + 800 $^{\circ}$C + 165 $^{\circ}$C/48h-$N_2$ @ 300 $^{\circ}$C & $1.83 \pm 0.01$ & $5.4 \pm 0.2$ & 11$\pm9$ &  9.28$\pm0.04$ \\
 	&& $\sim 25 \mu$m EP & $1.86 \pm 0.02$ & $1.5 \pm 0.3$ & 478$\pm 9$ & 9.30$\pm0.05$ \\
		 & & EP + 800 $^{\circ}$C + 155 $^{\circ}$C/48h-$N_2$ @ 155 $^{\circ}$C & $1.82 \pm 0.02$ & $2.9 \pm 0.2$ & 11$\pm 2$ & 9.30$\pm0.04$ \\
RDT-06	 &1.3  &  EP + 800 $^{\circ}$C + 155 $^{\circ}$C/48h-$N_2$ @ 250 $^{\circ}$C &$1.83 \pm 0.02$ & $1.3 \pm 0.2$ & 18 $\pm 2$ & 9.29$\pm0.05$ \\
		 & &  EP + 800 $^{\circ}$C + 165 $^{\circ}$C/48h-$N_2$ @ 300 $^{\circ}$C &$1.85 \pm 0.02$ & $1.0\pm 0.3$ & 5$\pm 1$ & 9.28$\pm0.03$ \\
		 & &  EP + 800 $^{\circ}$C + 165 $^{\circ}$C/24h-$N_2$ @ 300 $^{\circ}$C& $1.85 \pm 0.02$ & $1.9 \pm 0.2$ & 14 $\pm 1$ & 9.30$\pm0.05$ \\ 
RDL-02 &1.5  &  $\sim$ 25 $\mu$m EP & $1.82 \pm 0.02$ & $0.5 \pm 0.1$ & 364$\pm9$ & 9.28$\pm0.04$ \\
		&	&  EP + 800 $^{\circ}$C + 165 $^{\circ}$C/48h-$N_2$ @ 300 $^{\circ}$C & $1.85 \pm 0.02$ & $0.9 \pm 0.1$ &  4$\pm 1$ & 9.30$\pm0.03$ \\
 &  & $\sim$ 90 $\mu$m BCP+$\sim$ 40 $\mu$m EP &-- & -- & -- & -- \\
		& &+ 120 $^{\circ}C$/48 h & -- & --& -- & -- \\
FH3C		  & 3.0&  + 800 $^{\circ}$C + 165 $^{\circ}$C/48h-$N_2$ @ 300 $^{\circ}$C & $1.84 \pm 0.02$ & $9.5 \pm 0.8$ & 5$\pm1$ & 9.30$\pm0.05$ \\
		& &  +10 $\mu$m BCP+ 800 $^{\circ}$C + 165 $^{\circ}$C/48h-$N_2$ @ 300 $^{\circ}$C & $1.85 \pm 0.02$ & $2.7 \pm 0.4$ & 7$\pm1$ & 9.27$\pm0.05$ \\
\hline
\end{tabular}
\end{table*}

\section{Discussion}
In the past, LTB at 100 - 150 $^{\circ}C$ under UHV has been the standard practice for the final preparation of SRF cavities in order to recover from the high field Q-slope. Different models, related to oxide layer modification, oxygen diffusion, hydrogen segregation, and micro-structural modifications near the surface have been proposed to explain the high-field Q-slope and the LTB effect \cite{GigiAPL,gigiPRAB2010,marc}.

In our present study, the introduction of nitrogen during the low temperature baking showed improvement on $Q_0$, as well as the elimination of high field Q-slope, similar to that obtained by LTB in UHV environment. The improvement on $Q_0$ was clearly evident when the $N_2$ was injected at higher temperature $\sim$ 250-300 $^{\circ}C$ during the cooldown of the cavity from 800 $^\circ$C.  Our earlier studies showed that the presence of NbN$_{1-x}$O$_x$ layer between the bulk niobium and top most Nb$_2$O$_5$ layer may be responsible for the high $Q_0$ \cite{Dhakalinf}. The electronic properties of such layer and their influence on the electronic density of states of the adjacent superconducting Nb might explain the difference in the RF performance of current cavities compared to those which were subjected to the standard UHV baking. The surface modifications due to thermal treatments are also evident from the results of studies using x-ray photoelectron spectroscopy, magnetization and ac susceptibility measurements of samples treated with the cavities \cite{Dhakalinf}.

The RF measurements and sample surface analysis showed that the heat treatment at lower temperature significantly alters the RF surfaces, mostly driving the superconducting Nb towards the dirty limit, where the electronic mean free path is less than the superconducting coherence length within the RF penetration depth \cite{Dhakalinf}. Recent work on heat treatment of SRF cavities at medium temperature (300 - 400 $^{\circ}$C) also showed the increase in quality factor \cite{sam,Zhou,Ito,Eric}, associated to the diffusion of oxygen from the oxide layer. 

\begin{figure}[ht!]
\includegraphics*[width=85mm]{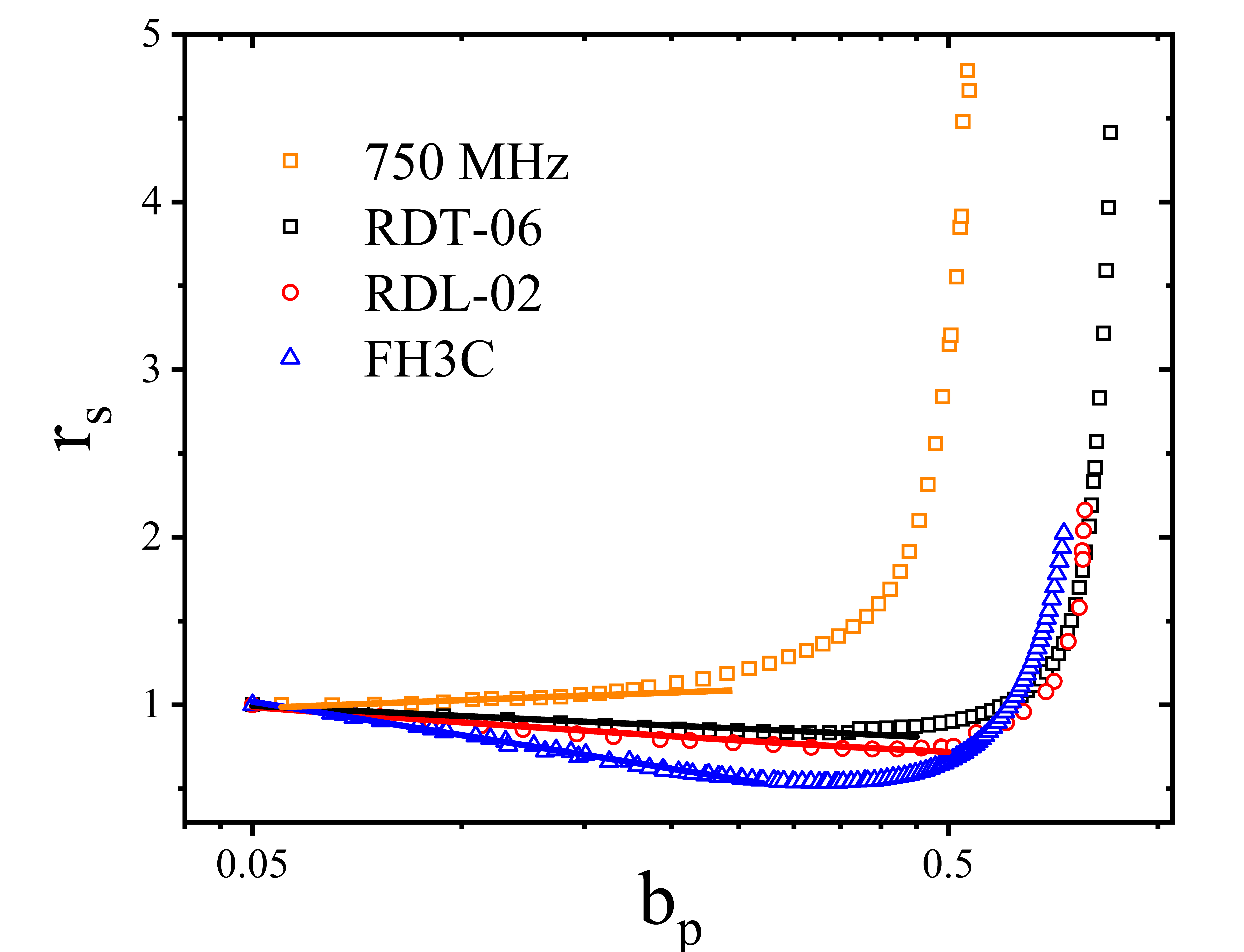}
\caption{\label{fig:ALL_Rs} Normalized $R_{BCS}$ for all cavities at 2.0 K after "N-infusion". The solid lines represent a linear fit of the data at low field.}
\end{figure}

\begin{figure}[ht!]
\includegraphics*[width=85mm]{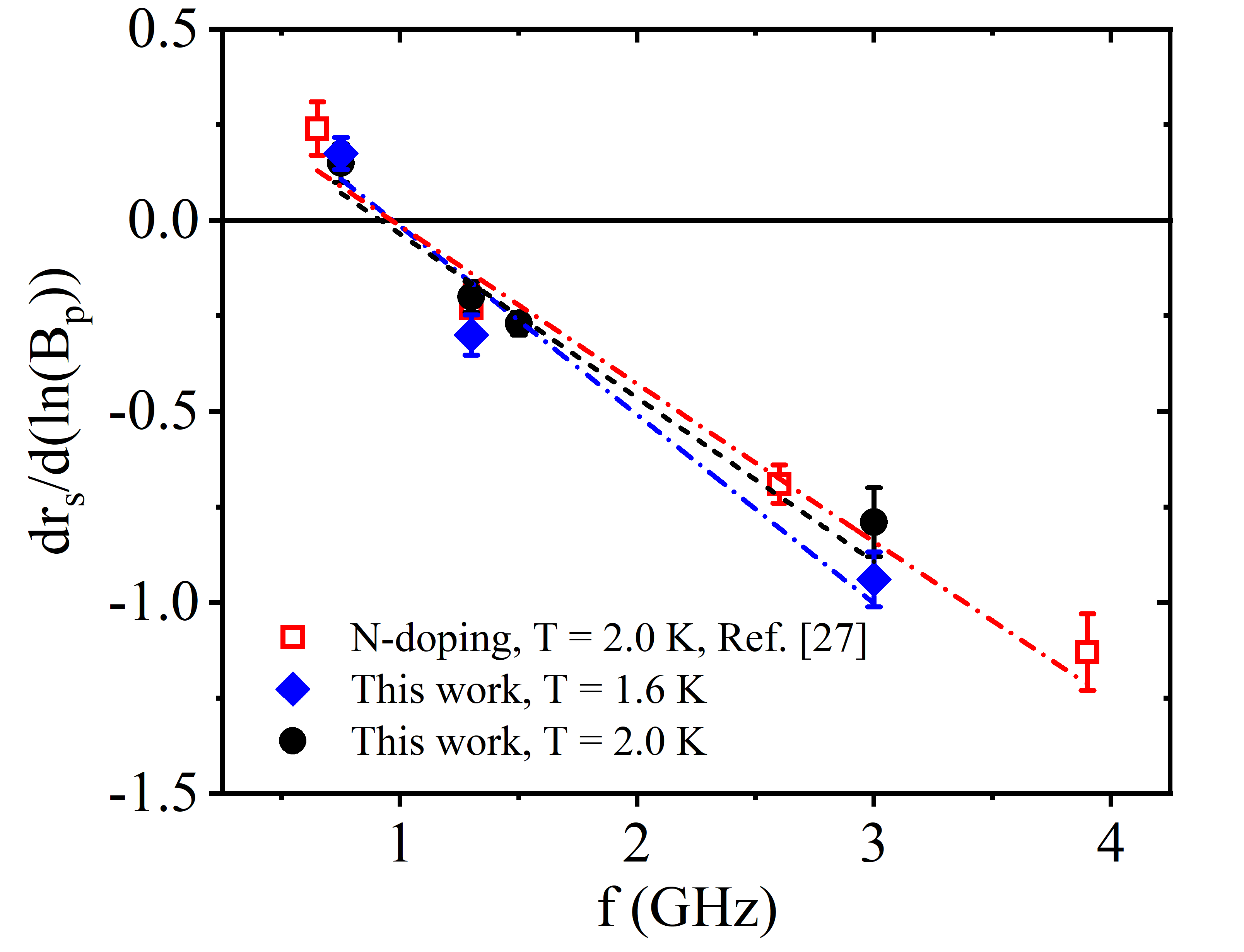}
\caption{\label{fig:slope} The slope of normalized BCS resistance with respect to ln$B_p$ from Fig. \ref{fig:ALL_Rs} for this study at 1.6~K and 2~K and from Ref. \cite{martina} at 2.0 K.}
\end{figure}

To get an insight into the complex dependencies of the surface resistance on RF field, frequency and temperature in the N-infused cavities, 
we summarized some essential features of the observed $R_s(B_p,f,T)$ in Figures \ref{fig:ALL_Rs} and \ref{fig:slope}.  Shown in 
Figure \ref{fig:ALL_Rs} is a normalized BCS part of the surface resistance $r_s = R_{BCS}(B_p)/R_{BCS}(\text{10 mT})$ as a function 
of the reduced RF field $b_p=B_p/B_c$ for all investigated cavities subjected to the same recipe of nitrogen treatment: 800~$^{\circ}$C/3h followed by 165~$^{\circ}$C/48 h with nitrogen injection into the furnace at ~$\sim 300$~$^{\circ}$C. One can see that the BCS temperature-dependent part of $R_s(B_p,f,T)$ decreases approximately linearly with $\text{ln}(B_p/B_c)$ up to $B_p\sim 50$~mT, similar to that was observed on N-doped Nb cavities~\cite{GigiAPL14}.  However, the negative slope of $r_s[\text{ln}(b_p)]$ in smaller cavities resonating at higher frequency changes sign and becomes positive in larger cavities resonating at lower frequencies. Similar behavior was also reported after the N-doping treatment in which the nitrogen is injected at 800~$^{\circ}$C \cite{martina}. Figure~\ref{fig:slope} shows the slope of $r_s[\text{ln}(B_p)]$ extracted from Fig. \ref{fig:ALL_Rs} for different cavity frequencies and two temperatures of 1.6 and 2K, along with the data from Ref.~\cite{martina}. The slope changes from positive (Q-slope) to negative (Q-rise) at a crossover frequency $f_0\simeq 0.95$~GHz. One can see that changing temperature from 1.6 to 2 K does not affect much either the slope and the frequency $f_0$ which turn our to be close to those observed on N-doped cavities treated at 800~$^{\circ}$C \cite{martina}.

\begin{figure}[htb]
\includegraphics*[width=85mm]{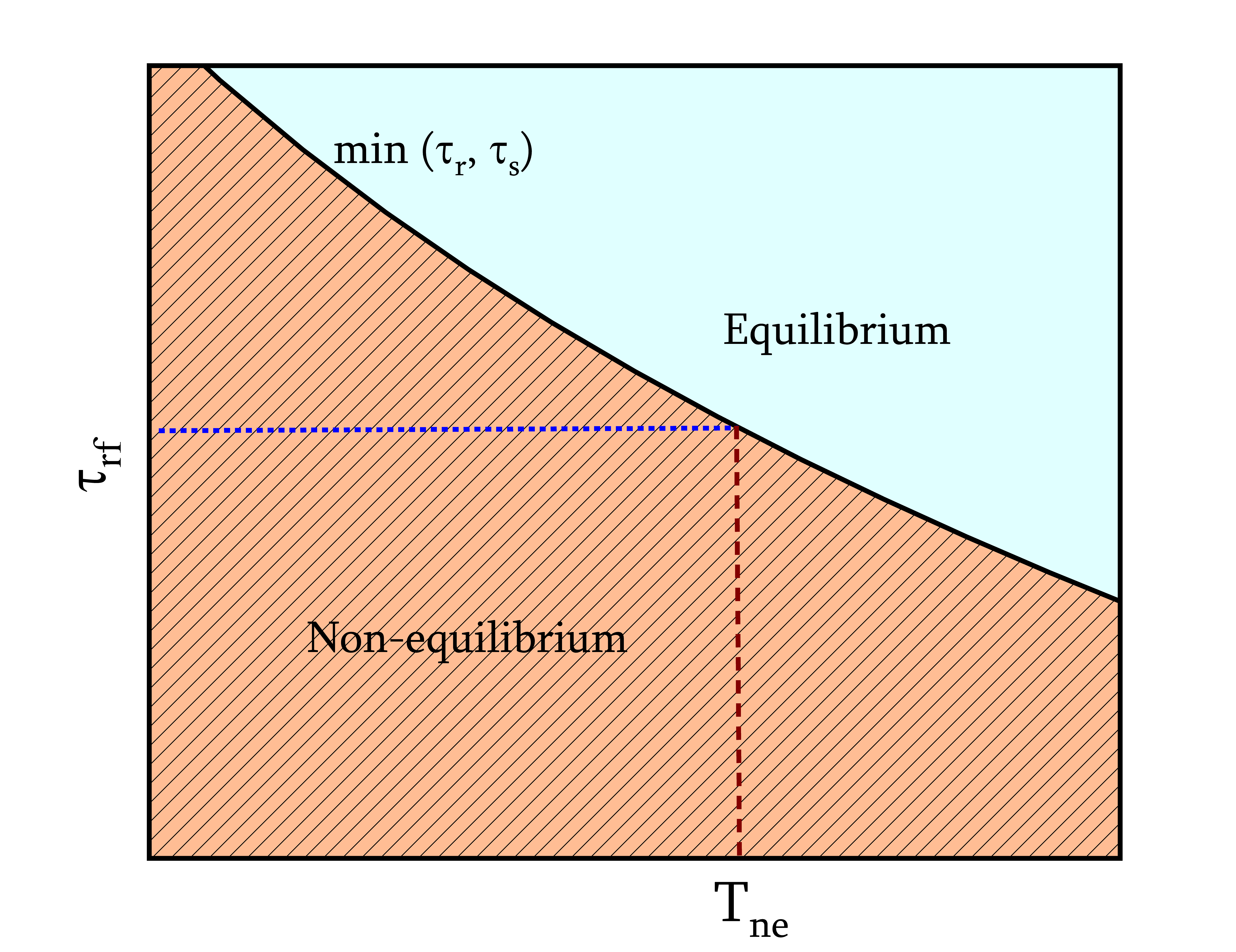}
\caption{\label{fig:eq_noneq} The RF period ($\tau_{rf}$) and temperature ($T_{ne}$) dependence criteria to observe the transition from equilibrium to non-equilibrium effects. The non-equilibrium effect is expected when $\tau_{rf}<\text{min}(\tau_r, \tau_s)$}
\end{figure}

It has been shown that the Q-rise can result from the well-established effect of broadening of the quasiparticle density of states by RF current \cite{AlexPRL, AlexSUST}. In this model the Q-rise can be amplified by non-equilibrium effects which occur if the RF period $1/f$ becomes smaller than the quasi-particles energy relaxation time $\tau_s(T)$ due to inelastic scattering on phonons ~\cite{kaplan,kopnin}. 
\begin{equation}
\tau_s(T)=\tau_0(T_c/T)^{7/2},
\label{taus}
\end{equation}
where $\tau_0$ is a material constant. If $f\tau_s\gtrsim 1$ the RF power absorbed by superconducting quasiparticles cannot be fully transferred to phonons during the rf period, so the distribution function of quasiparticles becomes non equilibrium and different from the Fermi distribution. Because $\tau_s(T)$ increases strongly as $T$ decreases, nonequilibrium effects become essential at lower temperatures, as shown in Fig. \ref{fig:eq_noneq}. 
For Nb at 2~K, Eq. (\ref{taus}) yields $\tau_s~2 \times 10^{-8}$~s and  $f\tau_s=20$ at 1GHz ~\cite{AlexPRL,kuboAlex}, suggesting that the quality factors of Nb cavities at intermediate and high rf fields can be affected by nonequilibrium quasiparticles driven by rf field. However, the clean-limit Eq. (\ref{taus}) may not give reliable numbers for the actual $\tau_s$ as scattering on impurities at the surface, proximity coupled NbO layer, subgap and two-level states can significantly shorten  $\tau_s$ \cite{alexsust23}. Yet the model Ref. \cite{AlexPRL} can describe the Q-rise observed on Nb cavities after nitrogen and titanium diffusion and low temperature baked cavities in nitrogen environment \cite{Dhakalinf}. 

\begin{figure}[htb]
\includegraphics*[width=85mm]{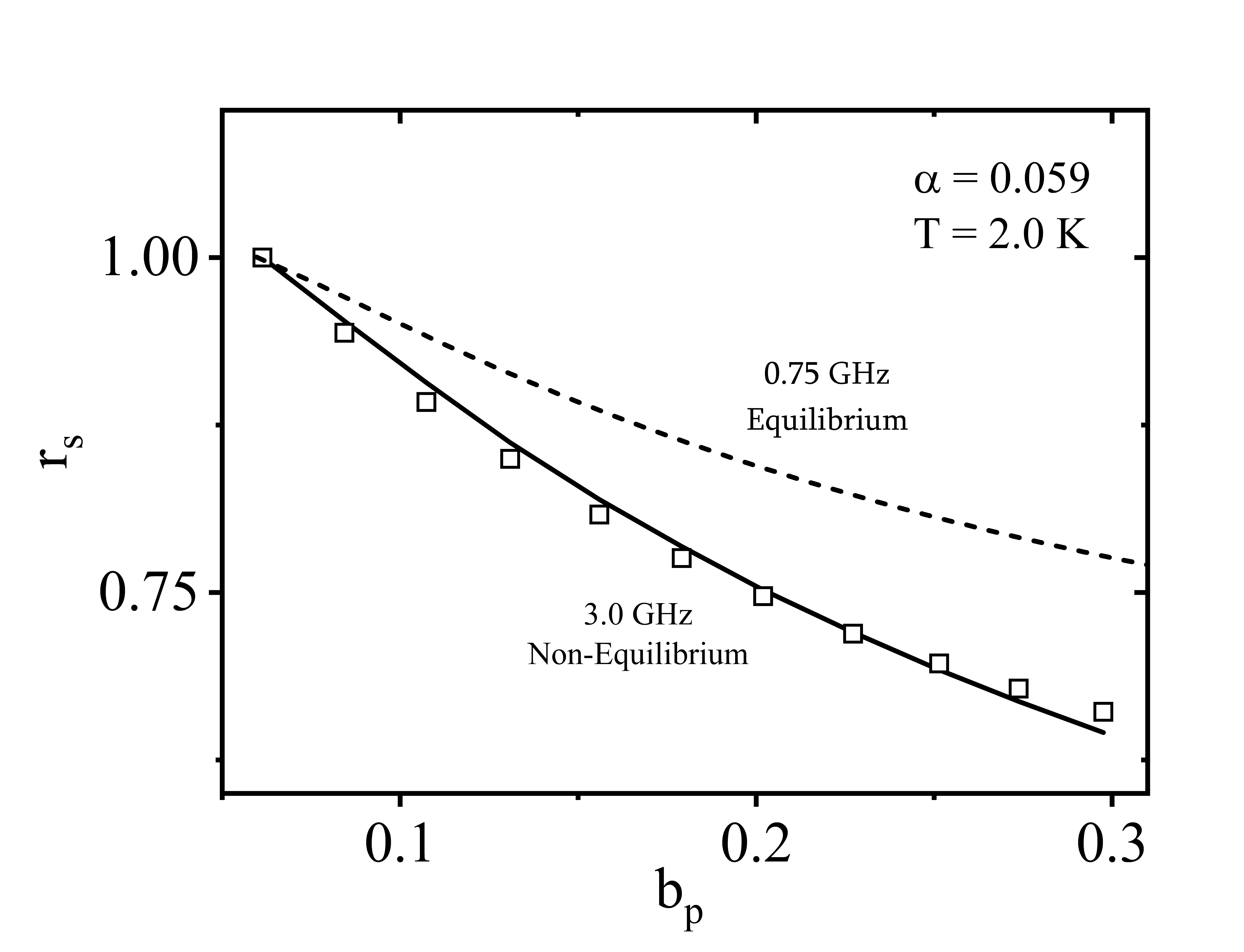}
\caption{\label{fig:eq_noneq_f} {Fit of $r_s$ vs. $b_p$ data (open square) at 2.0 K using the model of Refs. \cite{AlexSUST} for $f = 3.0$~GHz} using nonequilibirum distribution of quasiparticles. Also shown is the calculated $r_s$ as a function of $b_p$ for $f= 0.75$~GHz using the equilibrium distribution function. Here $\alpha$ is an overheating parameter defined in Ref. \cite{AlexPRL} and $B_c=180$ mT \cite{DeSorbo}.}
\end{figure}

\begin{figure}[h]
\includegraphics*[width=85mm]{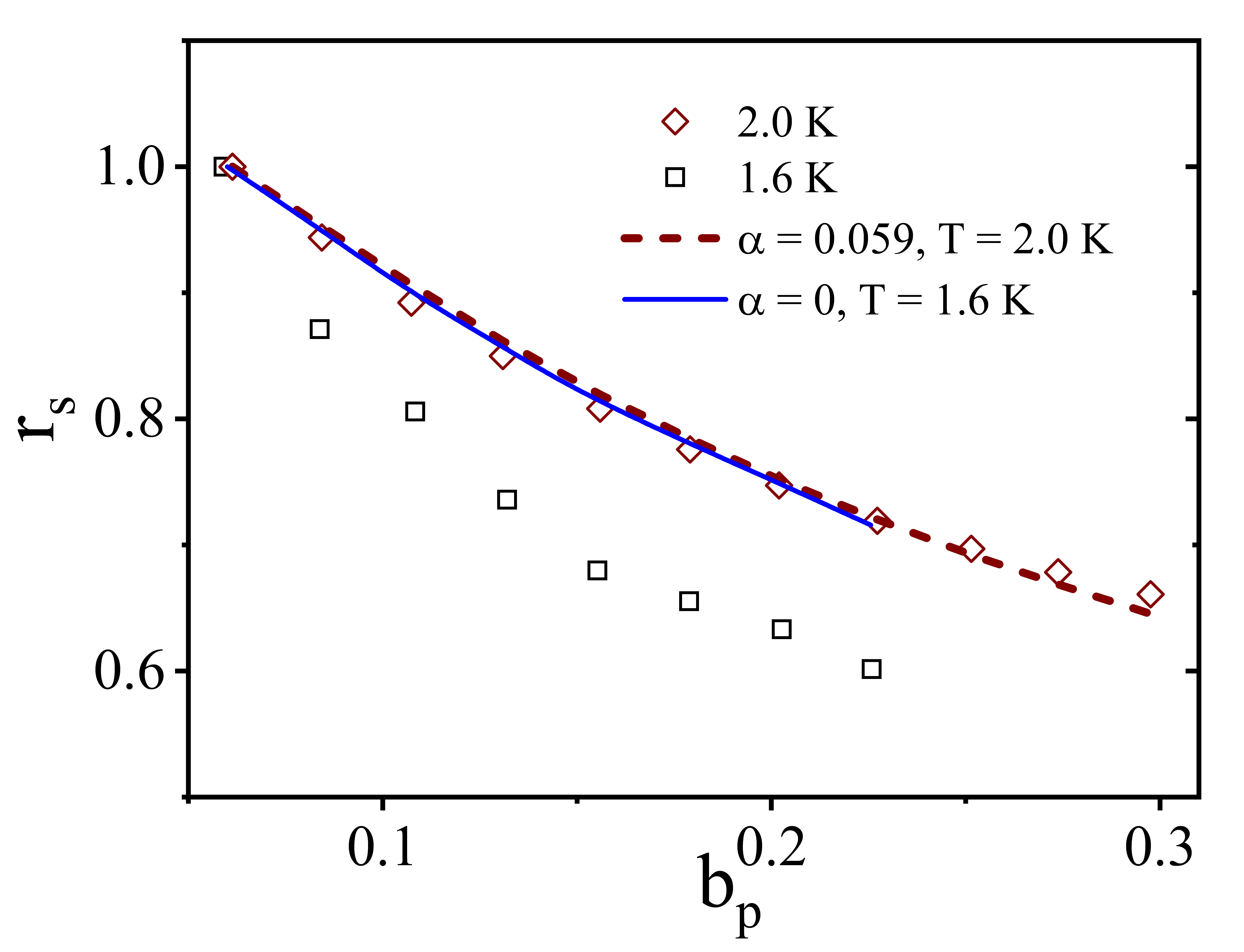}
\caption{\label{fig:RBCS_3GHz} $r_s$ vs. $b_p$ at different temperatures at 3~GHz. The dashed lines is the fit to the model of Ref.~\cite{AlexPRL} with the overheating parameters $\alpha = 0.059$ and $\alpha = 0$ at 2~K and 1.6~K, respectively and $B_c=180$ mT \cite{DeSorbo}.}
\end{figure}

Shown in Fig. \ref{fig:eq_noneq_f} is an example of the fit of $r_s(b_p)$ at 2~K measured in the 3~GHz cavity to the model with a nonequilibrium distribution function proposed in Ref. \cite{AlexPRL}, as well as $r_s(b_p)$ calculated with the equilibrium distribution function at 2~K and 750~MHz. Here the model with a nonequilibrium distribution function does describe the negative Q-slope at 2K and 3GHz, whereas the equilibrium Fermi distribution yieds a weaker negative Q-slope which is not sufficient to explain the experimental data. Yet the negative slope of $r_s[\text{ln}(b_p)]$ at lower temperature 1.6 K shown in Fig. \ref{fig:RBCS_3GHz} further increases by $\sim 40 \%$ as compared to $r_s[\text{ln}(b_p)]$ at 2 K. Similar results were reported in "N-doped" cavities at 1.3 GHz \cite{dhakalIPAC15}.  The model nonequilibrium distribution of Ref. \cite{AlexPRL} is no longer sufficient to fit the data at 1.6 K. In this case the field-dependent $R_s(B_a,f,T)$ at low temperatures should be calculated by solving numerically the full dynamic equations of nonequilibrium superconductivity including kinetic equations for the distribution function \cite{kopnin}.    

Now we turn to the transition from the Q-rise at higher frequencies $f>f_0$ to the Q-drop at lower frequencies $f<f_0$ The Q-rise caused by the broadening of the gap peaks in the idealized BCS density of states by RF current occurs even for the equilibrium distribution function ~\cite{AlexPRL, AlexSUST} (see Fig. \ref{fig:eq_noneq_f}). However, such Q-rise can be reversed to the Q-drop by magnetic impurities, subgap quasiparticle states or proximity-coupled metallic suboxides at the surface ~\cite{kuboAlex}. These effects can be ameliorated by the N-infusion, 
as was revealed by scanning tunneling microscopy of coupons cut off from the nitrogen-treated Nb cavities  \cite{EricPRA, Ericfrontier}. 
So one possibility of the Q slope reverse at $f<f_0$ may be due to the transition from the equilibrium state at $f\tau(T)\lesssim 1$ to a nonequilibrium state at $f\tau(T)\gtrsim 1$. Here the Q-slope at $f<f_0$ becomes positive because of the combined effect of subgap states and proximity-coupled suboxide which, however, are not sufficient to reverse the Q-rise in the nonequilibrium state at $f\tau\gtrsim 1$ in which the Q-rise is stronger, as shown in Fig. \ref{fig:eq_noneq_f}. If this in indeed the case, the crossover frequency $f_0\sim \tau^{-1}\propto (T/T_c)^{7/2}$ would decrease strongly with temperature, $f_0(1.6 K)=0.39f_0(2.1 K)$. This scenario is inconsistent with our data shown in Fig. \ref{fig:slope}, where $f_0(T)$ is practically the same at 1.6 K and 2K. Moreover, $f_0(T)$ measured on our cavities turns out to be the same as $f_0$ measured at FNAL on Nb cavity subject to different treatments~\cite{martina}. Such insensitivity of $f_0$ to the temperature and the cavity treatments appears inconsistent with the key role of nonequilibrium effects in the sign change of the Q-slope, since the inelastic scattering and recombination times of quasiparticles are strongly temperature dependent and sensitive to the concentration of impurities and the surface oxide structure \cite{alexsust23}.     

\begin{figure}[ht!]
    \centering
    \includegraphics*[width=85mm]{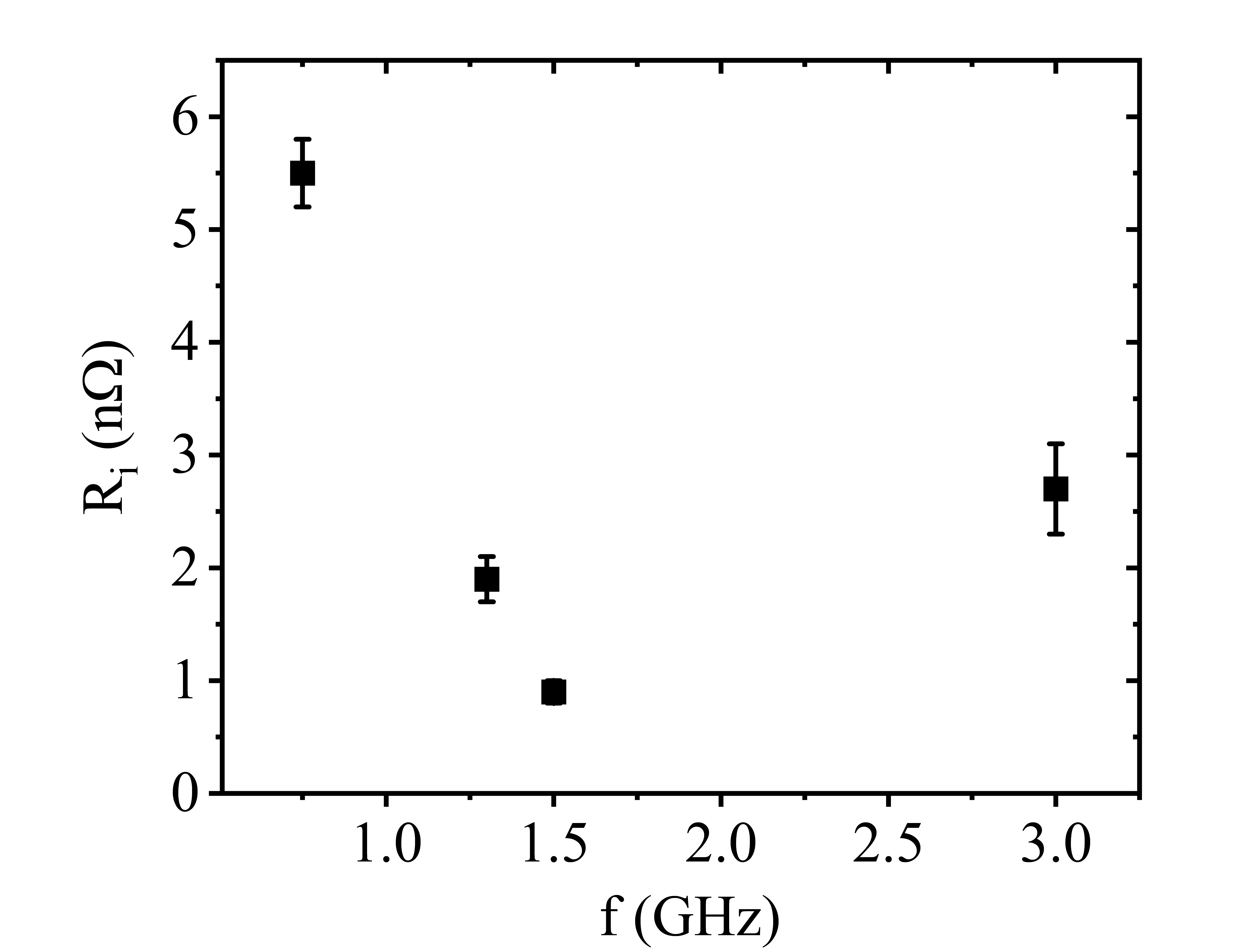}
    \caption{Residual resistance as a function of frequency of cavities of different sizes after the same surface treatment listed in Table \ref{table2}}
    \label{fig:Ri}
\end{figure}

Another possibility of the sign change of the Q slope at $f<f_0$ is that the Q-rise coming from the BCS surface resistance is masked by field-dependent contributions to the residual resistance.  To explore this scenario we measured $R_i$ on cavities resonating at different frequencies and subject to the same materials treatment.  The results plotted in Fig.~\ref{fig:Ri} show that $R_i$ increases sharply for the 750 MHz cavity as compared to the 3~GHz cavity, contrary to the conventional decrease of $R_i$ as the frequency decreases observed on Nb cavities \cite{halbritter84, GigiIEEE} and expected from $R_i$ resulting from trapped vortices \cite{AlexSUST,cornell,AlexManula}. Yet much lower $R_i$-values were observed on the same cavities after different treatments, as shown in Table~\ref{table2}. The data shown in  Fig.~\ref{fig:Ri} suggest the presence or a higher density of non-superconducting precipitates or trapped vortices in the 750~MHz cavities after the N-infusion treatment. Given the same treatment for these cavities, one can assume that their materials defect and the surface suboxide structures are not very different so the difference in $R_i$ comes from much larger density of trapped vortices in 1.3 GHz and 0.75 GHz cavities. Vortex trapping can be very dependent on the cavity size as the temperature inhomogeneities facilitating flux trapping upon cooling through $T_c$ are more pronounced in bigger cavities with smaller $f$. Furthermore, the mean free paths $l\simeq 4-11$ nm of these cavities listed in Table II are shorter than the clean-limit coherence length $\xi_0\simeq 40$ nm. As a result, the lower critical field $H_{c1}\simeq (l/\xi_0)H_{x1}^{clean}$ and the vortex core size $\xi\simeq \sqrt{l\xi_0}$ are reduced, facilitating penetration of vortices upon cooling through $T_c$ and enhancing flux trapping as smaller vortex cores enhance flux pinning \cite{AlexManula}.   

The observed sign reversal of the  Q-slope at $f<f_0$ can be understood if $R_i$ in larger cavities with smaller $f$ is dominated by trapped vortices. Indeed, large-scale numerical simulations of curvilinear vortices in a random pinning potential of materials defects \cite{AlexManula} have shown that $R_i(B_a)$ increases with $B_a$ due to depinning of vortices by the rf current, which can counter the field reduction of $R_{BCS}(B_a,f,T)$. Moreover, the masking effect of trapped vortices becomes more pronounced as the frequency decreases because  $R_i(B_a,f)\propto \sqrt{f}$  for weak pinning and  $R_i(B_a,f)\propto f$ for strong pinning \cite{AlexManula,cornell} decrease with $f$ much weaker than  $R_{BCS}(B_a,f,T)\propto f^2$. As a result, $R_i$ becomes dominant at low frequencies, resulting in the positive slope $dR_s/d\ln B_a$. 

The above consideration implies that $f$ is the only control parameter, whereas the density of trapped flux, the pinning defect structure, and inhomogeneities of superconducting properties are the same in all cavities. These stringent conditions are naturally satisfied in coaxial half-wave cavities in which $f$ can be varied by exciting different cavity modes \cite{delayen, Kolb, rautsrf23, KolbmidT},  but hard to implement in Nb elliptical cavities of different sizes, even if they have been treated the same way. A clear sign that $R_i$ in cavities of different sizes is affected by extrinsic factors is the breakdown of the continuous decrease of $R_i(f)$ with $f$ due to a sharp upturn in $R_i(f)$ at 1.3 and 0.75 GHz shown in Fig. \ref{fig:Ri}.  This is indicative of either higher density of trapped flux, as was discussed above, or uncontrolled inhomogeneities of superconducting properties, different densities of impurities or pinning nanoprecipitates, or  variations of the thicknesses of the suboxide NbO layers. For instance, the mean free path extracted from the BCS fit of $R_{BCS}(T)$ in cavities of different sizes listed in Table \ref{table2} varies from 4 nm to 11 nm.  These issues show challenges of separation of intrinsic and extrinsic contributions to $R_i(B_a,f)$ and extraction of a true $R_{BCS}(B_a,f,T)$ from the measured $R_s(B_a,f,T)$ to establish whether the sign change of  $dR_s/d\ln B_a$ at $f<f_0$ is related to the kinetics of nonequilibrium quasiparticles or results from extrinsic effects of trapped flux. This procedure becomes particularly ambiguous at lower frequencies and temperatures at which $R_i$ becomes larger than $R_{BCS}$ and a significant contribution to $R_i$ comes from quasiparticle subgap states and proximity-coupled metallic Nb suboxides \cite{kuboAlex}.

\section{Summary}
Improvement in the quality factor of SRF Nb cavities was observed after annealing at 800 $^{\circ}$C/3 h in vacuum followed by baking at 120 - 165 $^{\circ}$C  in low partial pressure of nitrogen inside a furnace  compared to the traditional 120 $^{\circ}$C bake in UHV. The improvement in $Q_0$ with Q-rise was observed only when the gas was injected in the furnace at 300~$^{\circ}$C during the cooldown from the 800~$^{\circ}$C. The negative slope of the normalized temperature dependent surface resistance as a function of ln($b_p$) increased with increasing frequency, with a crossover frequency $f_0\approx 0.95$ GHz below which the Q-slope becomes positive. The observed increase of negative Q-slope with decreasing temperature is qualitatively consistent with a temperature-dependent transition of quasiparticles driven by strong rf field to a nonequilibrium state controlled by electron-phonon energy relaxation. Further theoretical development may be needed to fully understand the frequency dependence of the Q-rise, yet the totality of our experimental data suggests that the observed change in the Q slope below 0.95 GHz may result from extrinsic factors such as  trapped flux, masking the underlying decrease of the quasiparticle surface resistance with $B_p$. Thus, unraveling the frequency dependence of the non-linear surface resistance may require multi-mode half-wave resonators, which minimize variability related to material processing, cavity size and trapped flux \cite{delayen, Kolb, rautsrf23, KolbmidT}. 

High quality factor at high accelerating gradient in SRF cavities would be of great interest for lowering the cryogenic heat load of high-energy accelerators such as the proposed Linear Collider \cite{ILC}. Processes that result in a reduced surface resistance in Nb SRF cavities may be relevant for the development of SRF cavities in superconductor-based quantum computers \cite{alex, alexK} as well as dark matter research \cite{backes}.

\section{Acknowledgments}
We would like to acknowledge Jefferson Lab technical staff members for the cavity surface processing and cryogenic support. We would also like to thank E. Lechner for writing a Matlab program to do least-squares fitting with the model described in Refs.~\cite{AlexPRL, AlexSUST} . This is authored by Jefferson Science Associates, LLC under U.S. DOE Contract No. DE-AC05-06OR23177. The work of Gurevich was supported by DOE under grant DE-SC
100387–020.

\end{document}